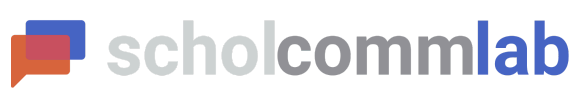

# A dataset of article processing charges from 14 scholarly publishers, 2019–2025

Lisa Matthias[1,2,*], Diego Chavarro[2,4], Eric Schares[2,3], Juan Pablo Alperin[2,4], Margaret Rose[2,5], Molly Frost[2,5], Flavia Camargo[2,5], Jonas Höfting[1,2], Leigh-Ann Butler[2,6], Nina Schönfelder[7] & Stefanie Haustein[2,5]

[1] Berlin School of Library and Information Science, Humboldt-Universität zu Berlin (Germany)
[2] Scholarly Communications Lab, Ottawa/Vancouver (Canada)
[3] University Library, Iowa State University, Ames (USA)
[4] Simon Fraser University, Vancouver (Canada)
[5] School of Information Studies, University of Ottawa, Ottawa (Canada)
[6] University of Ottawa Library, Ottawa (Canada)
[7] University Library, Bielefeld University, Bielefeld (Germany)

[*] *Corresponding author: l.a.matthia@gmail.com*

## Keywords

Article processing charges, scholarly publishing, open access, hybrid OA, gold OA, scholarly journals

## Abstract

This paper introduces a dataset of APCs produced from the price lists of 14 large scholarly publishers between 2019 and 2025. APC price lists were downloaded from publisher websites each year as well as via Wayback Machine snapshots to retrieve fees per journal per year. The dataset includes journal metadata, APC collection method, and annual APC price list information in several currencies (USD, EUR, GBP, CHF, JPY, CAD, AUD) for 12,540 unique journals and 69,856 journal-year combinations. The dataset was generated to allow for more precise analysis of APCs and can support library collection development and scientometric analysis estimating APCs paid in gold and hybrid OA journals.

## 1. Introduction

This paper introduces an open dataset (Matthias et al. 2026) of article processing charge (APC) prices compiled from publisher price lists, designed to support scientometric research, analyses of the scholarly publishing market, and library collections management. This dataset builds on and extends Butler et al.'s (2024) dataset by capturing annual APC prices beyond 2023 and including eight additional publishers.

In the following sections, we describe the methods used to collect, process, and clean the APC list prices to generate a coherent and clean dataset that comprises annual open access (OA) fees for gold (fully OA) and hybrid (subscription journals offering OA for individual articles) journals published by American Chemical Society (ACS), Cambridge University Press (CUP),

De Gruyter, EDP, Elsevier, Frontiers, IEEE, IOP, MDPI, Oxford University Press (OUP), PLOS, Sage, Springer Nature and Wiley from 2019 to 2025. We then summarize and discuss the dataset at a high level by presenting preliminary analysis and observations. We conclude by addressing potential uses of this open dataset.

# 2. Data and methods

## 2.1 Data sources

The dataset combines and standardizes data from the APC price lists of 14 large publishers. The dataset includes APC prices for 12,540 unique journals and 69,856 data points, so-called journal-year combinations, spanning seven years (2019-2025). In addition to the 37 publisher price lists used in the first version of the dataset, a further 59 APC lists and websites as well as 21 journal lists were incorporated and cleaned to produce one coherent and reusable dataset.

APC prices were captured from several sources and in various formats, as summarized in Table 1. Data collection followed two main approaches: in some cases, price lists were regularly downloaded directly from publisher websites by two of the authors (ES, NS); in other cases, archived snapshots of publisher websites containing the price lists were retrieved via the Wayback Machine by manually identifying and downloading relevant URLs. Price lists varied in format across publishers, ranging from downloadable PDFs and structured XLSX files to HTML pages on publisher websites, and typically included information such as ISSNs, journal titles, OA status, APC list prices, and currencies. Where price lists were not available on publisher websites in a downloadable format, APCs were either collected manually or scraped from individual journal web pages and collated into one XLSX file for import.

**Table 1**

*Overview of price lists, data sources, formats, and collectors*

| Publisher | Data sources | Format | Collectors |
|---|---|---|---|
| ACS | Publisher website; Wayback machine | HTML | DC; LM; MR |
| CUP | Publisher website; Wayback machine; APC list | XLSX | DC; LM |
| De Gruyter | Wayback machine; APC list | HTML | LM; NS |
| EDP | Wayback machine | HTML | LM; MF |
| Elsevier | Publisher website; Wayback machine; APC list | PDF, XLSX | DC; FC; LM; MF; MR; NS |
| Frontiers | Publisher website; Wayback machine | HTML | ES; LM; NS |
| IEEE | Publisher website; Wayback machine; APC list | PDF, XLSX | DC; LM; MR |

| IOP | Publisher website; Wayback machine; APC list | XLSX | DC; LM |
|---|---|---|---|
| MDPI | Publisher website; Wayback machine | HTML | ES; LM |
| OUP | Publisher website; Wayback machine; APC list | XLSX | DC; LM; NS |
| PLOS | Publisher website | HTML | LM; NS |
| Sage | Publisher website; Wayback machine; APC list | PDF; XLSX | DC; LM |
| Springer Nature | Publisher website; Wayback machine; APC list | PDF; XLSX | DC; FC; JH; LM; MR; NS |
| Wiley | Publisher website; APC list | XLSX | LM; MF; NS |

Sage's gold OA portfolio required a different approach, as no archived snapshots of its APC price list were available for 2019 and 2021–2024—only gold OA journal lists could be retrieved. We cross-referenced these lists with Crawford's gold Open Access collections (2020, 2022, 2023, 2024, 2025) to compile a complete set of gold OA journals for those years, yielding 1,039 journal-year combinations. APCs came from Crawford where available (n = 836), from manual checks of individual journal pages otherwise (n = 177), and from Morrison et al.'s (2021) dataset for any remaining gaps (n = 26).

Collecting IEEE's hybrid OA data for 2019 also required a different approach, as no archived price list was available. We began with IEEE's 2022 hybrid journal list, added 2022 gold OA journals not yet gold in 2019, and removed any journal that did not exist in 2019. For each remaining journal, we searched the Wayback Machine: where an APC appeared in an archived snapshot, it was recorded directly; where snapshots mentioned no OA or APCs and no 2019 OA content could be found, the journal was classified as subscription-only and removed. For the rest, we consulted archived 2019 issues, since some journals include author instructions within the journal itself.

Similarly to the first version of the dataset (Butler et al. 2024), we aimed to select price lists published or archived around June of each year. This was achieved for the majority of cases—44 out of 82 archived snapshots fell within one month of June, and 66 within two months—though in 12 instances the closest available snapshot fell further from June due to irregular publication or archiving schedules. For the remaining two, only the year of publication was known, with no specific month recorded. Collecting price lists around June of each year was chosen primarily for consistency across publishers and years. However, a mid-year snapshot also equally balances potential over- and underestimation of fees paid in the first versus second half of the year. This approach differs from Delta Think, a commercial service that retrieves publisher price lists each January to distribute aggregated APC data to paying subscribers (Pollock and Staines 2024). As opposed to Delta Think's offering, our dataset is openly available.

## 2.2 Selection of files, data cleaning, and metadata enrichment

The dataset was constructed through a combination of manual data entry and algorithmic parsing of structured data. Manual entry was coordinated by one of the authors (LM), who led a group of researchers in interpreting and entering data in a standardized format. The process involved multiple team meetings in which ambiguous cases were discussed and shared data entry conventions were established. For instance, where journals listed different APCs for society members and non-members, we consistently selected the price for non-members. The algorithms developed for parsing followed the same rules as agreed upon in these team meetings, ensuring consistency across both manual and automated entries.

Algorithmic parsing was used for the following publishers and years: Cambridge University Press (2019–2025), Elsevier (2024, 2025), IEEE (2020–2025), IOP (2019, 2021–2025), OUP (2022–2025), Sage (2020, 2021), and Springer Nature (2024–2025). Different R scripts were developed to ingest each list, extracting journal and APC information, standardizing it to the dataset's schema (Table 2, Section 2.2.6), and deriving any missing currencies using a fixed conversion priority (USD → GBP → EUR → AUD → JPY). While all scripts followed the same general process, the specific characteristics of the different files, between and within publishers, required individual adaptation of the scripts. Differences in file format (Excel vs. HTML), column naming conventions, and field availability (e.g., ISSN) meant that each script required tailored extraction and parsing logic. Some publishers also presented additional complexity (e.g., IEEE and OUP, for example, required specific handling of member and non-member pricing splits), and comment generation rules varied across publishers.

We then merged all records into a single dataset. The merging script standardizes journal and publisher names, resolves missing information, and removes duplicates, prioritizing manual entries over those obtained through automated parsing. Currency conversion was then applied to express all APCs in seven currencies (USD, EUR, GBP, AUD, JPY, CAD, CHF). Following the merge, we carried out a series of manual checks on the combined dataset, including verification of missing or anomalous APC values, removal of closed access journals, and identification of potential duplicates not captured by the script, among other quality checks described in more detail below. This procedure yielded v2 of this dataset, which is openly available for reuse as Matthias et al. (2026) on the ScholCommLab's Dataverse.

Figure 1 shows whether data were collected for a given publisher and year, though it does not imply complete APC information for all journals within a publisher's portfolio for those years. For most publishers, we have both gold and hybrid OA data across all seven years; Frontiers, MDPI, and PLOS are fully gold OA publishers, so the dataset contains gold OA records only for them. Coverage is incomplete for a few publishers: De Gruyter and OUP begin only in 2021, IOP has a gap in 2020, and Sage has only gold OA data in 2019 and only hybrid OA in 2025. In these cases, we could not find APC or journal lists to confidently establish the full portfolio for the missing years or OA type—though individual IOP (2020) and Sage gold OA (2025) records were identified through other price lists and retained in the dataset.

**Figure 1**

*APC data coverage by publisher and year*

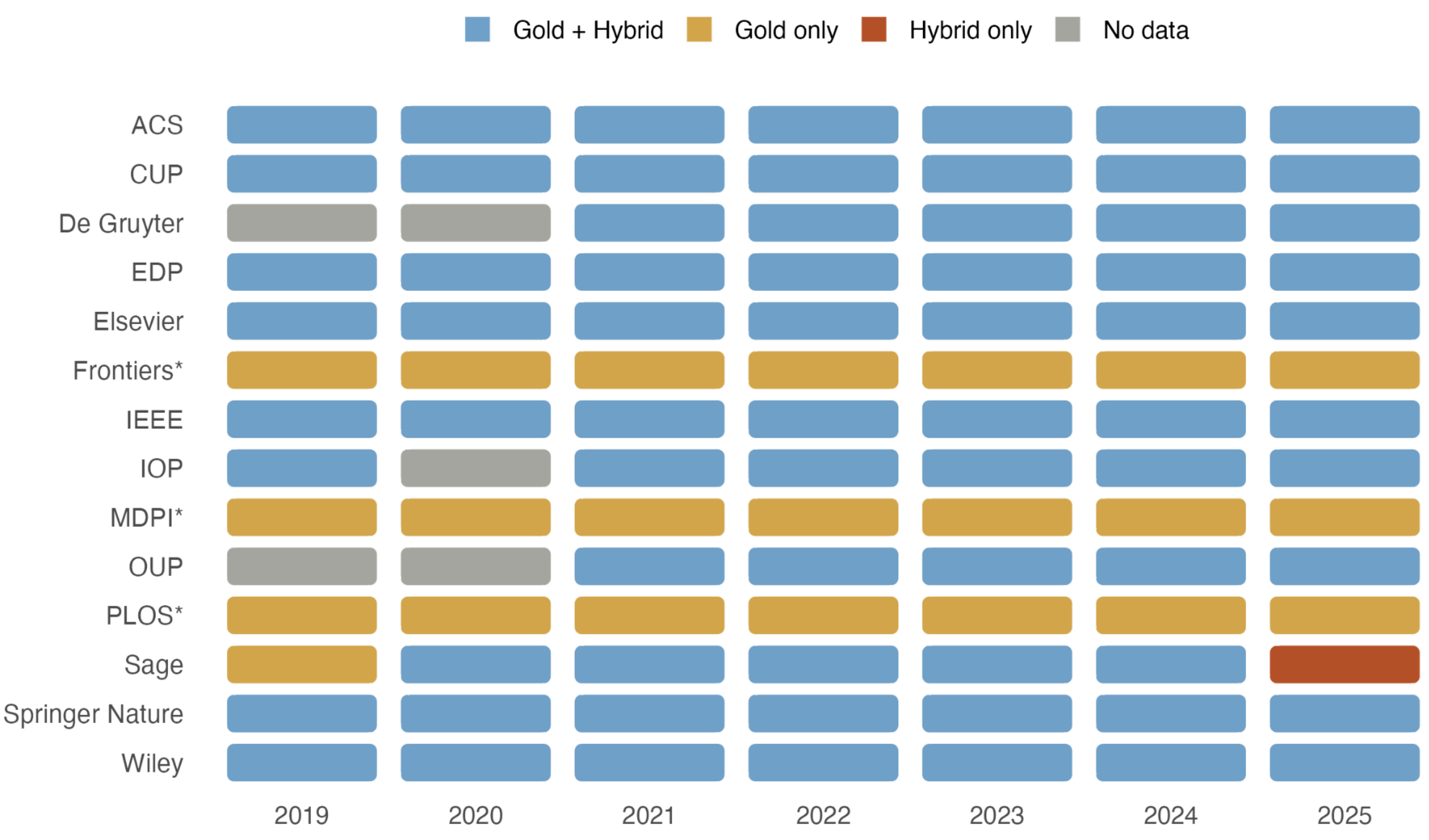

### 2.2.1 Unique identifiers

The dataset has two internal identifiers. Each APC record carries a unique numeric ID, ensuring every row is distinct. A separate journal-level ID links all records for the same journal across years, even where it appears under different names, publishers, or ISSNs—variations that arise from publisher switches or inconsistent listings. These were matched using a Python algorithm, and this ID should be used when counting unique journals. Both identifiers are internal only: they carry no meaning and correspond to no external identifier.

### 2.2.2 ISSNs

Most publisher price lists include at least one ISSN but do not always indicate whether it is the print (pISSN) or electronic (eISSN) version. The dataset therefore does not distinguish between them, instead using two columns (ISSN_1 and ISSN_2); the linking ISSN (ISSN-L), retrieved from the ISSN.org portal, is stored separately (ISSN_L). All ISSNs were standardized with a hyphen after the first four digits.

We validated ISSNs against the ISSN.org portal by comparing the registered title with the price-list title, scored using Jaro-Winkler fuzzy matching (Winkler 1990): scores above 80 were accepted automatically and those below 15 were flagged, with borderline cases reviewed using a local large language model (Ollama). Nearly all records (n = 12,032; 96%) were validated automatically; most of the remaining records (n = 357; 2.8%) were validated with the LLM; only six (0.05%) were rejected by the LLM and were checked manually. Of these, all but one were valid; the one confirmed error was an ISSN misassignment in the publisher's price list, where Sage's *Environmental Policy and Law* had been given the ISSN of the *Journal of Alzheimer's Disease*. The remaining 145 (1.16%) were fetch failures and were validated manually.

### 2.2.3 Changes in publisher

In some cases, the same journal was associated with more than one publisher, typically as a result of a journal switching publishers. Transfers were usually confirmed through announcements on journal, publisher, or society websites and noted in the comment column; transferred journals retain the same unique ID across all records. In 75 cases, the journal appeared in both publishers' price lists for the same year, producing two APC records for that year—in rare cases, across several years. For example, the *Journal of Interactive Marketing* (unique_id = 1799) appeared in both Elsevier's and Sage's lists from 2022–2024, though it transferred from Elsevier to Sage in 2022. To avoid duplicate entries for the same journal-year, we retained only the new publisher's entry from the year of transfer onward.

### 2.2.4 Journal title variations

Where a journal appeared under multiple title variants, including inconsistencies in diacritics, we standardized to the most frequent variant per journal ID.

### 2.2.5 Currency conversion

Publisher price lists vary in the currencies used, with many providing fees in multiple currencies. The dataset retains all originally provided values—USD, EUR, GBP, JPY, CHF, CAD, and AUD—and derives any missing currency by converting from an available one using annual average exchange. Daily exchange-rate observations were obtained from Yahoo Finance (Yahoo Finance 2026) and converted into annual average exchange rates by taking the arithmetic mean of available daily observations within each calendar year, using the quantmod R package (Ryan et al. 2026). Exchange rates are included in the dataset (Matthias et al., 2026). USD was the most common, provided directly for 94% of the 68,506 flat-rate journal-year combinations; the remaining 6% were converted from AUD, CHF, EUR, GBP, or JPY. For each currency, a column (e.g., apc_usd_originalORconverted) flags whether the value came directly from the publisher ("original") or was converted ("converted from CHF").

### 2.2.6 Machine readability

To ensure the dataset was machine-readable, fees with mixed data types were split across two columns. The apc column holds values expressible as a single number (flat rates, and fees such as rapid service or editorial processing charges); where a flat rate carried an additional charge (e.g., overlength or submission fees), the flat rate was recorded in the apc column and the extra charge in the apc_text column. Fees that cannot be expressed as a single number (e.g., page charges, scaled APCs) are recorded entirely in apc_text. Table 2 lists all variables with a description and example value.

**Table 2**

*Variables in the dataset*

| Variable name | Description | Example value |
|---|---|---|
| record_id | Unique ID per record. | 9580 |
| unique_id | Unique ID per journal. | 1782 |
| publisher | The publisher of the journal in the particular year. | Elsevier |
| issn_1 | The first ISSN of the journal, either print or electronic. | 0022-1759 |
| issn_2 | The second ISSN of the journal, either print or electronic, if provided. | 1872-7905 |
| issn_l | The linking ISSN (ISSN-L) of the journal, if provided. | 0022-1759 |
| journal | The proper title of the journal (normalized). | BURNS |
| oa_status | The open access status of the journal, as indicated by the publisher, journal website, or price list. | Hybrid OA |
| type_of_fee | The type of fee the article processing charge (APC) represents. Where no fee information could be found, the value is recorded as unknown. | flat rate |
| apc_*currency**<br>** for each of following currencies: USD, EUR, GBP, JPY, CHF, CAD, AUD* | The APC value as provided by the publisher or converted into each respective currency. | 2,950 |
| apc_*currency*_originalORconverted | Indicates whether the value came directly from the publisher's price list or was converted from another currency using an annual conversion rate. | original |
| apc_text | Additional APC information for entries whose type_of_fee is not a flat rate (e.g., page charges or scaled fees). | 300-900 EUR |
| apc_date | Date of collection of the price list, or of the Wayback Machine snapshot. | 2019-06-01 |
| apc_year | Year of the price list, or of the Wayback Machine snapshot. | 2019 |
| apc_source | Original source of price list. | Publisher website |
| collector | Name of person who collected the price list. | N. Schönfelder |
| comment | Further contextual or significant details. | Journal transferred from Sage to Wiley in 2019 |
| data_version | Version of the dataset release: 1 for records from the first release (Butler et al., 2024), 2 for records added in the second (Matthias et al., 2026). | 1 |

## 3. Descriptive statistics of the dataset

This section provides a statistical overview of the dataset, covering the number of unique journals per publisher and year and the distribution of APC list prices across publishers, years, and OA status. Table 3 shows the size differences among publisher portfolios and how they evolved between 2019 and 2025. Elsevier, Springer Nature, and Wiley consistently held the largest portfolios, growing to 2,985, 2,844, and 1,845 journals by 2025, respectively. Among fully gold OA publishers, MDPI grew most, from 205 journals in 2019 to 472 in 2025, and Frontiers from 61 to 228; PLOS remained the smallest, with 14 in 2025. Overall, journal-year records grew from 7,486 in 2019 to 11,286 in 2025, reflecting both genuine portfolio growth and improved coverage in the dataset. Although the dataset covers 12,540 unique journals in total, no single year reaches that number. This is because some journals appear in earlier years but drop out of later price lists—a journal may have moved to a publisher not included in this dataset, stopped offering OA publishing, or ceased publication.

**Table 3**

*Number of unique journal IDs per publisher, per year*

| Publisher | 2019 | 2020 | 2021 | 2022 | 2023 | 2024 | 2025 | Growth (%) |
|---|---|---|---|---|---|---|---|---|
| ACS | 59 | 75 | 77 | 79 | 84 | 87 | 89 | 51 |
| CUP | 326 | 370 | 380 | 387 | 395 | 419 | 438 | 34 |
| De Gruyter | NA | NA | 355 | 363 | 357 | 412 | 413 | 16 |
| EDP | 45 | 44 | 43 | 47 | 50 | 49 | 48 | 7 |
| Elsevier | 2,257 | 2,275 | 2,605 | 2,549 | 2,699 | 2,749 | 2,985 | 32 |
| Frontiers | 61 | 80 | 108 | 148 | 221 | 229 | 228 | 274 |
| IEEE | 135* | 212 | 212 | 196 | 195 | 210 | 216 | 60 |
| IOP | 71 | 1* | 79 | 87 | 85 | 87 | 97 | 37 |
| MDPI | 205 | 272 | 367 | 405 | 427 | 442 | 472 | 130 |
| OUP | NA | NA | 443 | 463 | 501 | 514 | 512 | 16 |
| PLOS | 7 | 7 | 12 | 12 | 12 | 14 | 14 | 100 |
| Sage | 213* | 1,148 | 1,155 | 1,067 | 1,159 | 1,223 | 1,085* | 409 |
| Springer Nature | 2,565 | 2,691 | 2,712 | 2,722 | 2,723 | 2,781 | 2,844 | 11 |
| Wiley | 1,542 | 1,609 | 1,657 | 1,665 | 1,888 | 1,893 | 1,845 | 20 |
| **All Publishers** | 7,486 | 8,784 | 10,205 | 10,190 | 10,796 | 11,108 | 11,286 | 51 |

*Note.* *Partial coverage only; no complete APC or journal list was available for this publisher in this year. Individual records may still be present where journals were identified through other price lists.

There were a total of 3,184 journal-year combinations with missing APC and/or OA status information, which were attributed to De Gruyter (n = 266), EDP (n = 89), Elsevier (n = 636), Frontiers (n = 52), IEEE (n = 244), IOP (n = 1), MDPI (n = 21), OUP (n = 1), Sage (n = 1,039), and Springer Nature (n = 835). After manual verification of each individual case using archived snapshots, 432 journal-year combinations in the final dataset retained missing APC values, and 23 lacked an OA status classification.

Of the 69,424 journal-year combinations with an APC value provided, 1,988 listed an APC of $0 (USD), indicating that authors were not required to pay a fee to publish OA in those journals. A zero APC may reflect a temporary waiver—for instance, for marketing purposes—or permanent waivers where publication costs are covered by a third party. While 1,983 of these records were gold OA journals, there were also five entries listed as a hybrid OA with an APC of $0, such the *International Journal of Obesity Supplements* published by Springer-Nature, *Chinese Physics C* published by IOP, which is part of the SCOP3 initiative, or *Dialogues on Climate Change* published by Sage under a sponsored S2O model.

For journal-year combinations with a flat-rate APC above $0, fees ranged from $100 for De Gruyter's hybrid journal *American Mineralogist* and Sage's gold journal *Tropical Conservation Science* to $12,690 for 39 Springer Nature hybrid journals (Table 4). The average APC across all publishers and years was $2,091 for gold OA and $3,278 for hybrid OA. Comparing price points by publisher, average gold OA APCs were lowest for EDP ($1,135) and highest for ACS ($3,955). Average hybrid OA fees varied more widely, ranging from $1,755 for EDP to $4,850 for ACS, with the three largest publishers—Elsevier ($3,121), Springer Nature ($3,324), and Wiley ($3,409)—clustering near the overall hybrid average.

Table 4 presents the distribution of flat-rate APC prices by publisher and OA status, including the number of journal-year combinations, minimum, median, maximum, interquartile range, mean, and coefficient of variation for all years. Beyond average prices, fees vary considerably across publishers and OA models. For many publishers, gold OA fees show greater relative variation than hybrid fees, reflected in higher coefficients of variation (CV%): Elsevier (48%), EDP (47%), and Wiley (41%) show the widest spread in gold OA pricing, while hybrid fees are often more uniform, with CVs as low as 5% (ACS), 10% (De Gruyter), and 11% (CUP). The interquartile range shows the same pattern—EDP's hybrid journals, for example, have a narrow IQR of $152, against $1,115 for their gold OA journals. Median values largely track the means.

**Table 4**

*APC price distribution by publisher and OA status*

| Publisher | OA status | N | Y | Min | Med | Max | IQR | Mean | CV |
|---|---|---|---|---|---|---|---|---|---|
| ACS | Gold | 66 | 7 | 750 | 3,750 | 5,000 | 1,500 | 3,955 | 31 |
| | Hybrid | 471 | 7 | 4,500 | 5,000 | 5,000 | 500 | 4,850 | 5 |
| CUP | Gold | 501 | 7 | 630 | 3,450 | 4,260 | 950 | 3,035 | 24 |
| | Hybrid | 2,166 | 7 | 511 | 3,255 | 4,920 | 185 | 3,191 | 11 |
| De Gruyter | Gold | 292 | 5 | 379 | 1,082 | 2,995 | 244 | 1,188 | 40 |
| | Hybrid | 1,385 | 5 | 100 | 2,366 | 2,712 | 326 | 2,350 | 10 |
| EDP | Gold | 103 | 7 | 226 | 1,250 | 1,921 | 1,115 | 1,135 | 47 |
| | Hybrid | 75 | 7 | 1,159 | 1,713 | 3,550 | 152 | 1,755 | 27 |
| Elsevier | Gold | 4,326 | 7 | 150 | 1,850 | 8,900 | 1,000 | 1,977 | 48 |
| | Hybrid | 13,328 | 7 | 150 | 3,000 | 11,400 | 970 | 3,121 | 32 |
| Frontiers | Gold | 1,040 | 7 | 950 | 2,125 | 3,801 | 1,050 | 2,276 | 30 |
| IEEE | Gold | 146 | 7 | 600 | 1,950 | 2,075 | 245 | 1,821 | 16 |
| | Hybrid | 409 | 7 | 1,750 | 2,345 | 2,995 | 800 | 2,513 | 16 |
| IOP | Gold | 117 | 6 | 675 | 2,395 | 3,750 | 515 | 2,250 | 22 |
| | Hybrid | 334 | 7 | 1,025 | 2,930 | 4,680 | 465 | 2,972 | 19 |
| MDPI | Gold | 2,536 | 7 | 304 | 1,207 | 3,499 | 739 | 1,517 | 39 |
| OUP | Gold | 569 | 5 | 260 | 2,474 | 4,937 | 1,096 | 2,451 | 34 |
| | Hybrid | 1,807 | 5 | 1,000 | 3,895 | 8,000 | 831 | 3,917 | 18 |
| PLOS | Gold | 78 | 7 | 1,595 | 2,575 | 6,460 | 650 | 2,902 | 37 |
| Sage | Gold | 965 | 6 | 100 | 1,575 | 4,470 | 800 | 1,634 | 36 |
| | Hybrid | 5,831 | 6 | 514 | 3,250 | 5,400 | 650 | 3,327 | 14 |
| Springer Nature | Gold | 3,533 | 7 | 518 | 2,390 | 7,990 | 1,020 | 2,455 | 37 |
| | Hybrid | 14,554 | 7 | 1,380 | 3,090 | 12,690 | 710 | 3,324 | 34 |
| Wiley | Gold | 2,286 | 7 | 600 | 2,310 | 6,730 | 1,123 | 2,277 | 41 |
| | Hybrid | 9,594 | 7 | 950 | 3,300 | 6,100 | 940 | 3,409 | 21 |
| **All Publishers** | Gold | 16,558 | 7 | 100 | 2,000 | 8,900 | 1,200 | 2,091 | 44 |
| | Hybrid | 49,954 | 7 | 100 | 3,150 | 12,690 | 870 | 3,278 | 28 |

*Note.* Flat-rate APCs > $0 only. N = number of journal-year combinations, Y = number of years with data, Min = minimum, Med = median, Max = maximum, IQR = interquartile range, CV = coefficient of variation.

## 4. Limitations

While this dataset aims to provide comprehensive APC coverage for some of the largest academic publishers, it has limitations. Coverage is constrained by the availability of APC information in publisher price lists and archived webpages—where data could not be retrieved, this is flagged in the dataset to ensure it is used appropriately. The dataset should therefore not be treated as an exhaustive record of all APCs charged by the included publishers, but rather as a systematic compilation of what was publicly available and retrievable at the time of collection.

## 5. Conclusion

This dataset provides annual APC list prices for 12,540 journals across 14 major scholarly publishers from 2019 to 2025, compiled into a single, openly available, machine-readable resource. By standardizing fees across publishers, currencies, and years, and by documenting OA status and fee type for each record, it lowers the barrier to APC-based research that would otherwise require assembling price lists from disparate and often archived sources.

The dataset supports several lines of reuse. For libraries and consortia, the dataset can be freely used to support analyses of OA investments, including Transformative / Read and Publish agreements, institutional trends, and/or collections development, to name a few. For bibliometric research, combining these list prices with publication data enables estimation of APC expenditure in gold and hybrid journals at the article, institutional, or national level. The longitudinal structure also enables study of how list prices developed over time and how they differ across publishers and OA models, while the record of publisher changes supports analysis of consolidation in the scholarly publishing market.

## Acknowledgements

The authors gratefully acknowledge Madelaine Hare, PhD candidate, Dalhousie University, for her early contributions on the initial version of the dataset and initial data paper.

## Author contributions

**Lisa Matthias:** Data curation, Methodology, Validation, Visualization, Writing – original draft, Writing – review & editing **Diego Chavarro:** Data curation, Methodology, Validation, Writing – original draft, Writing – review & editing **Eric Schares:** Data curation, Methodology, Validation, Writing – review & editing **Juan Pablo Alperin:** Conceptualization, Supervision **Margaret Rose:** Data curation, Methodology, Writing – review & editing **Molly Frost:** Data curation, Writing – review & editing **Flavia Camargo:** Data curation **Jonas Höfting:** Data curation **Leigh-Ann Butler:** Methodology, Writing – review & editing **Nina Schönfelder:** Data curation **Stefanie Haustein:** Conceptualization, Funding acquisition, Methodology, Supervision

## Competing interests

JPA and SH maintain a financial interest in pending litigation against four of the publishers discussed in this paper, which seeks recovery for allegedly unreasonable and unnecessary article processing charges.

## Funding information

This research was funded by the Volkswagen Foundation (Az.: 9C784).

## Data availability

The dataset described in this paper is openly available on the Harvard Dataverse at [https://doi.org/10.7910/DVN/AZ985C](https://doi.org/10.7910/DVN/AZ985C). The dataset is released under a Creative Commons CC0 public domain dedication. When using the dataset, please cite it as Matthias et al. (2026).